\documentclass[1p]{elsarticle}

\pdfoutput=1

\makeatletter
\def\ps@pprintTitle{%
 \let\@oddhead\@empty
 \let\@evenhead\@empty
 \def\@oddfoot{}%
 \let\@evenfoot\@oddfoot}
\makeatother

\usepackage{lineno,hyperref}

\usepackage{amssymb}
\usepackage{url}
\usepackage{amsfonts}
\usepackage{amsmath}
\usepackage{epsfig}
\usepackage{lineno}
\usepackage[usenames]{color}
\usepackage{hyperref}
\usepackage{ulem}
\usepackage{graphicx}
\begin{document}

\begin{frontmatter}

\title{Simulations of a multi-layer extended gating grid}

\author{J.D. Mulligan}
\address{Yale University, New Haven, CT, USA}
\ead{james.mulligan@yale.edu}

\begin{abstract}
A novel idea to control ion back-flow in time projection chambers is to use a multi-layer extended gating grid to capture back-flowing ions 
at the expense of live time and electron transparency. In this initial study, I perform simulations of a four-layer grid for the ALICE and 
STAR time projection chambers, using $\text{Ne}-\text{CO}_{2}\;(90-10)$ and $\text{Ar}-\text{CH}_{4}\;(90-10)$ gas mixtures, respectively. I report the 
live time and electron transparency for both 90\% and 99\% ion back-flow suppression. Additionally, for the ALICE configuration I study several effects:
using a mesh vs. wire-plane grid, including a magnetic field, and varying the over-voltage distribution in the gating region. 
For 90\% ion back-flow suppression, I achieve 75\% live time with 86\% electron transparency for ALICE, and 
95\% live time with 83\% electron transparency for STAR. 
\end{abstract}


\end{frontmatter}

%\linenumbers

\section{Introduction}

In high-rate gaseous time projection chambers (TPCs), ion back-flow (IBF) from the gas amplification region to the drift volume distorts the drift field,
deteriorating tracking and PID performance. A recent proposal to control IBF in TPCs is to use a multi-layer extended gating grid~\cite{wieman2015}. 
In comparison to a traditional gating grid, the extension of the grid with multiple layers allows a longer 
time for ions to drift through the gate, while still collecting the ions quickly.
The operating principle is that the gate remains transparent to electrons until the ion drift time exceeds the grid length (divided by the ion drift velocity), 
at which point the gate is closed and the ions are collected. Enhanced IBF suppression comes at the sacrifice of live time and electron transparency; 
for a given IBF tolerance, the design goal is to increase the live time fraction $A$ while maintaining sufficient electron transparency for reconstruction 
performance. Such a design could operate as a primary means of IBF suppression, or in cooperation with other elements such as Gas Electron Multipliers. 
Early work suggests that for a wire-plane gate, low-field regions between the wires prevent some ions from being captured quickly~\cite{wieman2015}. 
The detailed simulations presented below quantify this effect and serve as an initial study of the general feasibility of a multi-layer extended gating grid. 

\section{Simulation Configurations}

In order to study the performance of the grid in various TPC conditions, I simulate the gating region for two large gas TPCs: ALICE~\cite{alice} and STAR~\cite{star}. 
These TPCs use different gas mixtures (with significantly different ion mobilities) and different drift fields, which considerably impact the gating performance.

\begin{figure}[!h]
\centering{}
\includegraphics[scale=0.7]{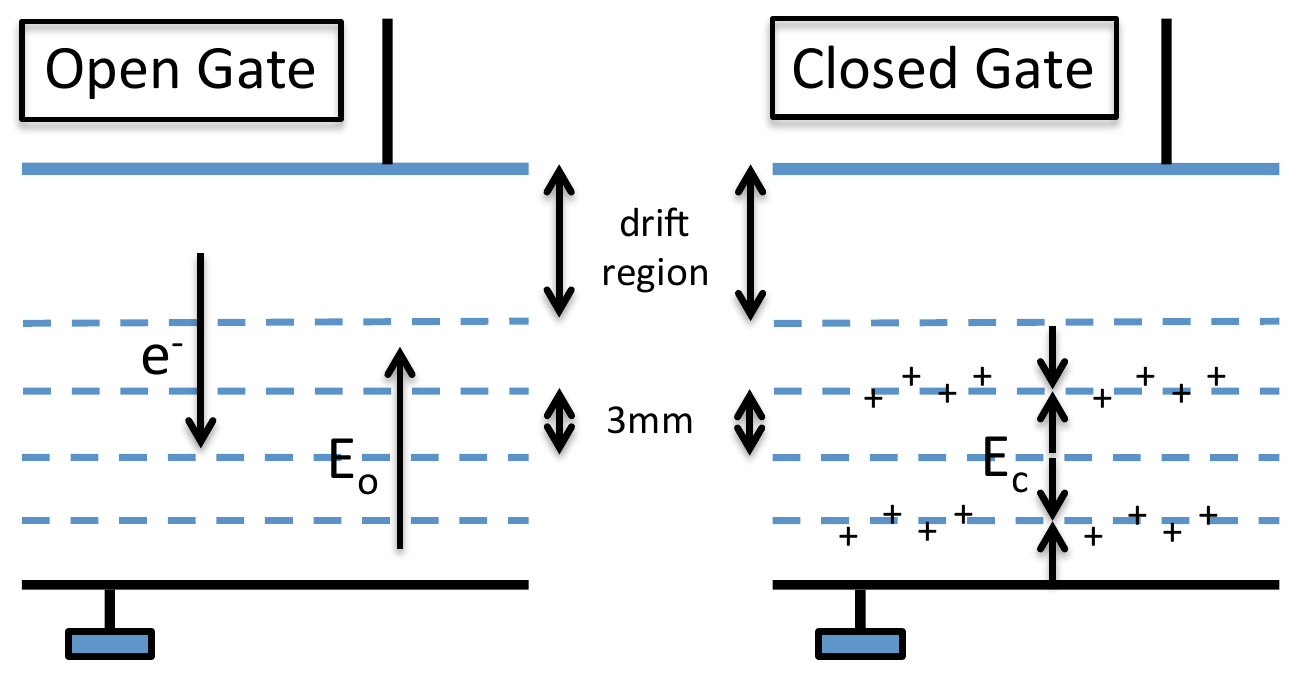}
\caption{Schematics of the open (left) and closed (right) gating configurations. In the open configuration, electrons pass through with small losses, while
positive ions back-drift through the gate. In the closed configuration, these back-drifted positive ions are collected on the first and third planes by $E_c$.}
\label{fig:schematic}
\end{figure}

In both TPC configurations, I consider a four-layer grid (Fig.~\ref{fig:schematic}), with the open field $E_o$ parallel/anti-parallel to the closed field $E_c$. 
The spacing between layers
is $3\;\text{mm}$, and the inter-layer wire spacing is $2\;\text{mm}$. I use a $3\;\text{mm}$ drift volume above the grid, 
and a grounded plane $3\;\text{mm}$ below the grid. The wire diameter is $100\;\mu\text{m}$.
The ions are collected on the first and third planes from the gas amplification region, with $E_c\approx2\;\text{kV/cm}$.

The fields are constructed using the finite element method 
software ANSYS~\cite{ansys}; the electron and ion drifts are simulated using Garfield++~\cite{garfield}. Collision-level tracking is performed for electrons 
(``microscopic tracking"), and a more coarse-grained Monte Carlo tracking is used for ions. Diffusion is included for both electrons and ions. It should
be noted that while Garfield++ can natively solve 2D fields, close examination revealed that the ANSYS solution is more accurate near the wires. 

\subsection{ALICE TPC}

For the ALICE TPC, I use a gas mixture of $\text{Ne}-\text{CO}_{2}\;(90-10)$, as configured for LHC Run 1. The drift field is 
$\approx0.4\;\text{kV/cm}$; a representative voltage switch required on the four gating planes (in volts) is: $(-600,0,-600,0)\leftrightarrow(-120,-240,-360,-480)$. The electron 
drift velocity in this mixture for the considered drift field is $2.73\;\text{cm}/\mu\text{s}$~\cite{tdr2014}. Binary ion mobilities from the 
literature are linearly extrapolated to low fields and combined for the gas mixture using 
Blanc's Law~\cite{viehland1995}\cite{kaneko1977}\cite{blum2008}. The dominant ion in this mixture is $\text{CO}_{2}^{+}$~\cite{veenhof}.

\begin{figure}[!t]
\centering{}
\includegraphics[scale=0.39]{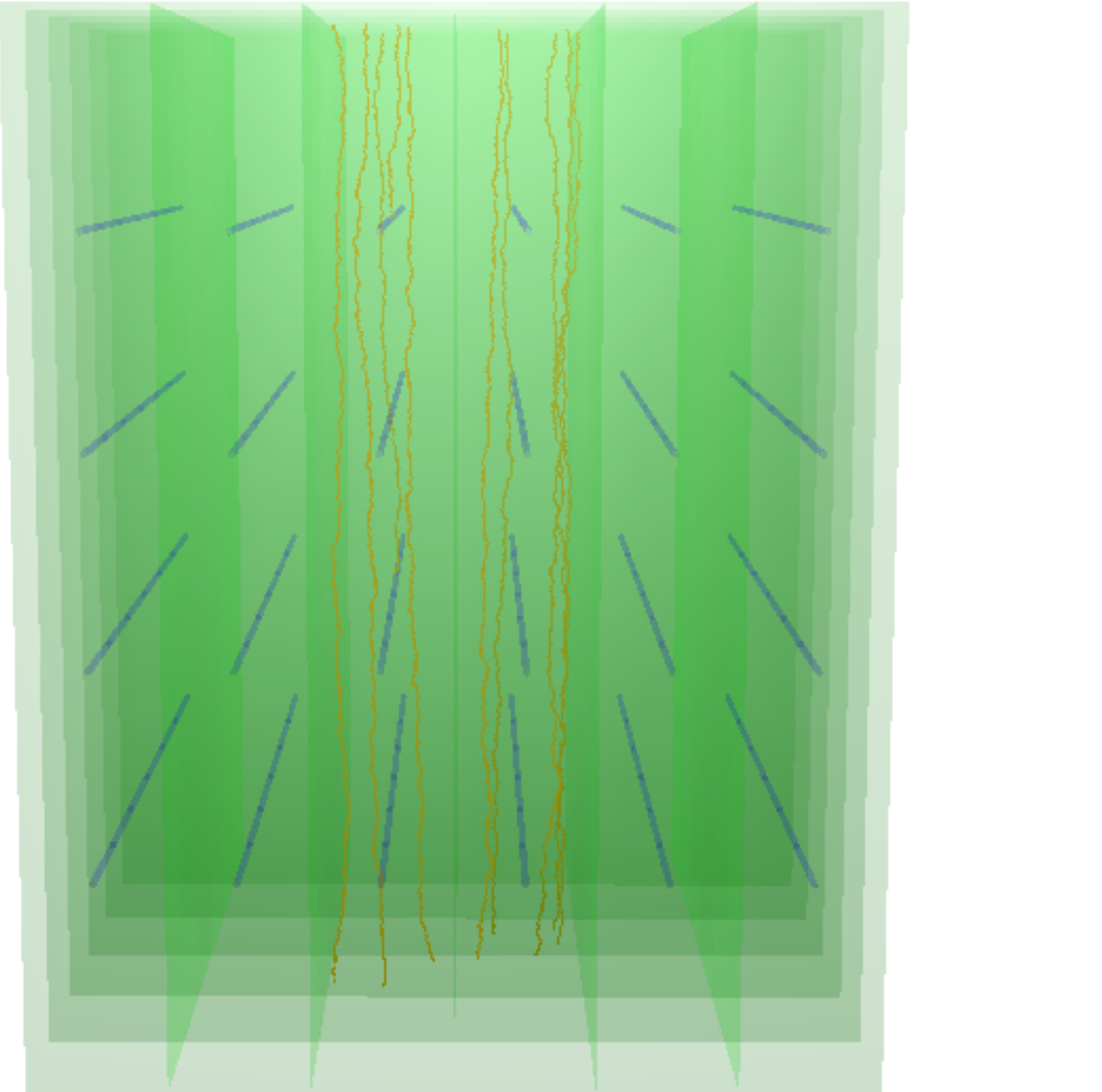}
\includegraphics[scale=0.39]{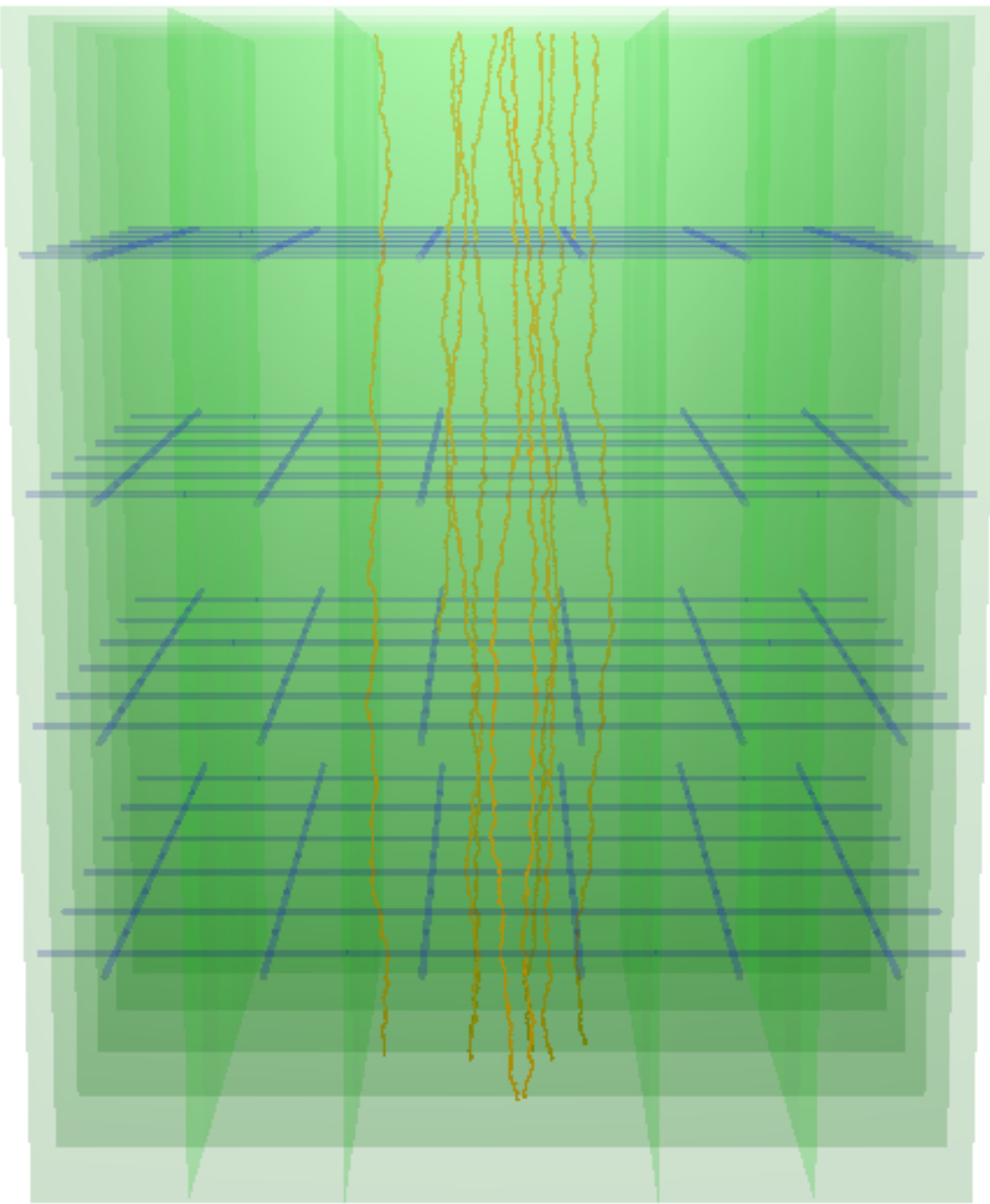}
\caption{Visualizations of the multi-layer extended gating grids in the open configuration, with electron drift lines traveling downward. Plotted
are $6\times6$ arrays of $2\text{mm}\times2\text{mm}\times15\text{mm}$ unit cells used for the simulation. Left: Wire-plane configuration.
Right: Mesh configuration. Mesh spacing in-plane is $2\;\text{mm}$.}
\label{fig:simSchematic}
\end{figure}

To study differences in ion collection time and electron transparency, I study separately a wire configuration and a mesh configuration (Fig.~\ref{fig:simSchematic}). 
The final finite element meshes contain approximately
$3\cdot10^{5}$ elements for the wire configuration unit cell, and $2\cdot10^{6}$ elements for the mesh configuration unit cell. These
correspond to an average element size of $56\;\mu\text{m}$ for the wire configuration, and $33\;\mu\text{m}$ for
the mesh configuration. However, adaptive meshing is employed, which creates finer elements near geometrical features. The final meshes
were examined for quality, and an informal convergence study was performed, in which iteratively refined meshes were produced, and 
the maximum field near the wires showed convergence to $<10\%$. 

\subsection{STAR TPC}

For the STAR TPC, I use a gas mixture of $\text{Ar}-\text{CH}_{4}\;(90-10)$. The drift field is approximately $140\;\text{V/cm}$. The electron drift
velocity in this mixture for the considered drift field is $5.45\;\text{cm}/\mu\text{s}$, and the ion mobility $1.6\;\text{cm}^2/\text{V}\cdot\text{s}$. 

Only a wire-plane configuration is simulated. The finite element mesh contains approximately $3\cdot10^{5}$ elements for the unit cell, as for the ALICE wire-plane mesh. 

\section{Simulation Results -- ALICE TPC}

\subsection{Electron Transparency}

\begin{figure}[!t]
\begin{centering}
\includegraphics[scale=0.7]{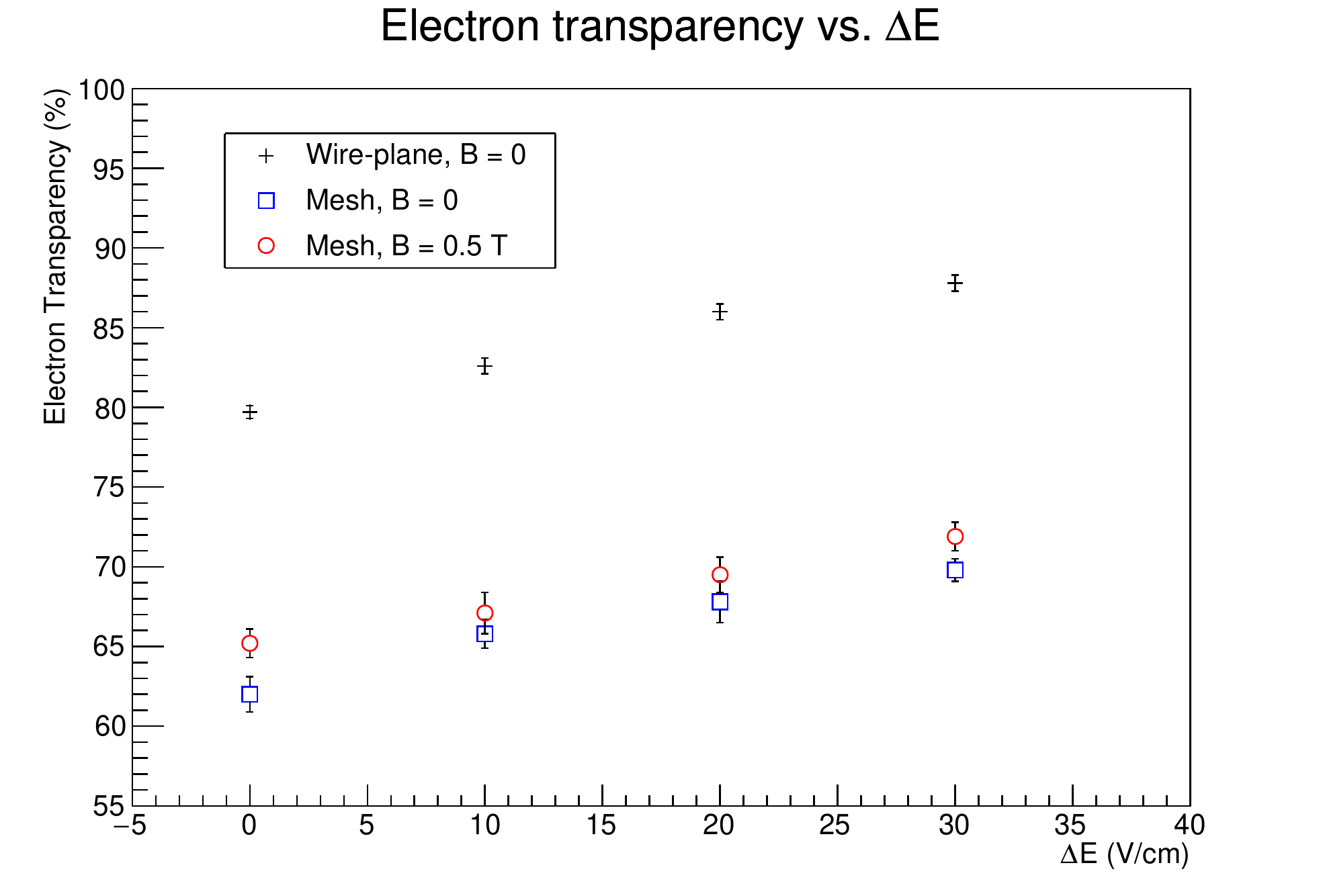}
\par\end{centering}
\caption{Electron transparency as a function of field incrementation $\Delta E$ at each layer of the grid, in the ALICE configuration. The error bars estimate the
statistical uncertainty. For the wire-plane configuration, each point corresponds to $2.5\cdot10^{4}$ electrons. For the mesh configuration,
each point corresponds to $10^{4}$ electrons. }
\label{fig:transparency}
\end{figure}

I measure electron transparency by randomly placing electrons at the top of the drift region, and measuring the fraction that pass
through the grid in the open configuration.
At each layer in the grid, I increment the field by a value $\Delta E$ in order to boost the transparency; increasing $\Delta E$ amounts
to putting negative charge on the wires, which repels drifting electrons. I use fixed over-voltages corresponding to 
$\Delta E=0,10,20,30\;\text{V/cm}$ across each plane, yielding average open gating fields of $E_o=400, 425, 450, 475\;\text{V/cm}$. 
Figure~\ref{fig:transparency} shows the results for both the wire-plane and mesh configurations. 

Additionally, I repeated electron transparency measurement in the mesh configuration with a magnetic field $B=0.5\;\text{T}$ parallel
to the electric field. This results in a slight increase in transparency (Fig.~\ref{fig:transparency}), which may be due to reduced transverse 
diffusion (from the $B$-field) outweighing $E\times B$ effects 
(which may deviate drifting electrons from the electric field lines into a wire); the cause remains to be investigated.

\begin{figure}[!t]
\begin{centering}
\includegraphics[scale=0.7]{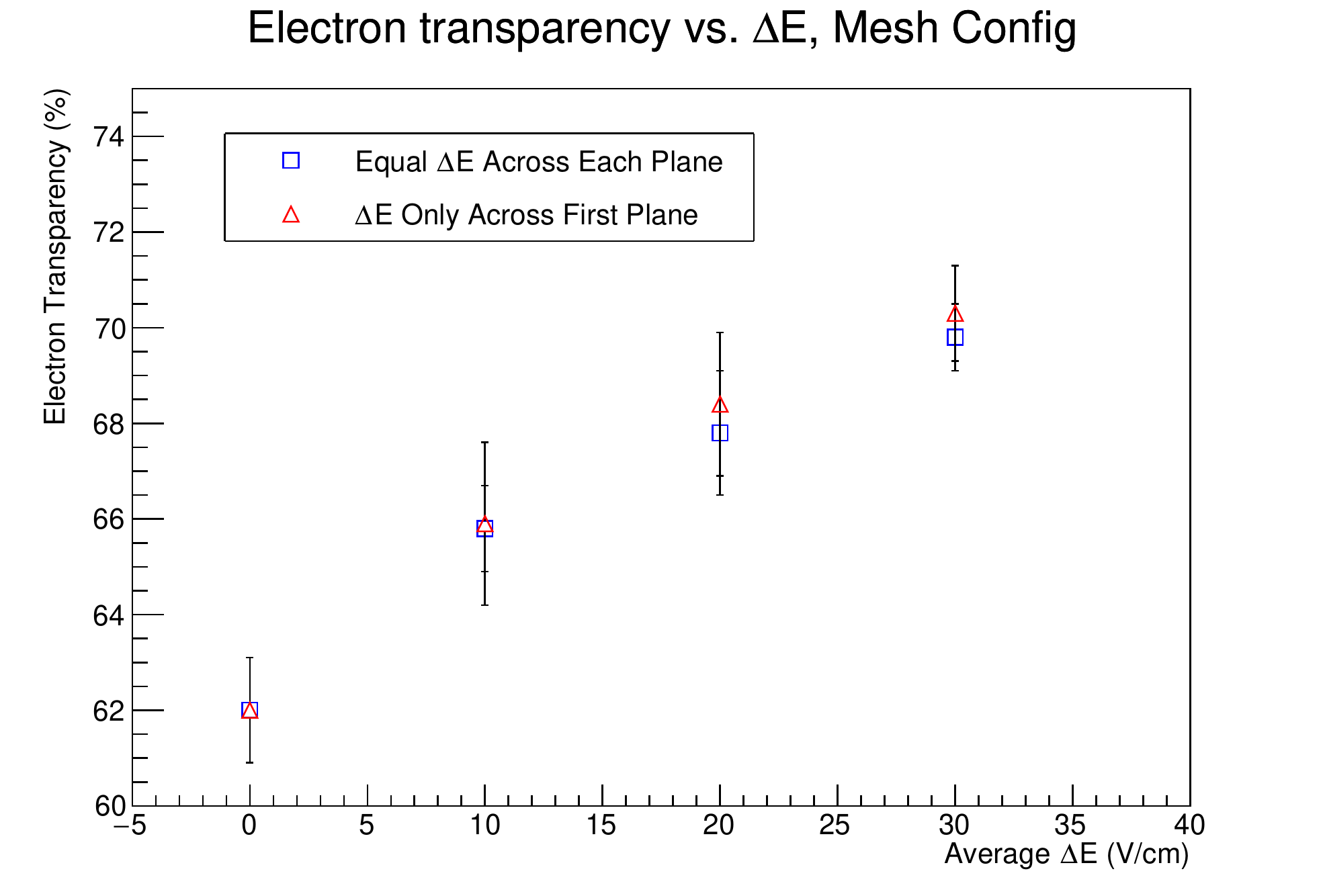}
\par\end{centering}
\centering{}\caption{Electron transparency as a function of average $\Delta E$ across
each layer of the grid, for two different overvoltage distributions, in the ALICE configuration.
Each point corresponds to $10^{4}$ electrons; the error bars estimate
the statistical uncertainty.}
\label{fig:overvoltage}
\end{figure}

Next, I examined the over-voltage distribution to determine if there is an optimal way to distribute $\Delta E$ over different planes,
rather than fixing it to be constant across each layer. A comparison of fixed $\Delta E$ over each plane against having nonzero $\Delta E$
only across the first plane shows little difference (Fig.~\ref{fig:overvoltage}). In the latter case, fewer electrons are captured on the first layer, but more are captured
in subsequent layers. This suggests that if more layers are added to the grid, the fixed $\Delta E$ configuration is better.

\subsection{Live Time and Ion Collection}

Following {[}1{]}, the live time fraction $A$ of the gating grid can be written \[ A=\frac{T_{active}}{T_{cycle}}=\frac{T_{o}-T_{e}}{T_{o}+T_{c}}, \]
where $T_{o}$ is the open time, $T_{c}$ is the closed time, and $T_{e}$ is the time for an electron to drift the length of the chamber.
For a gating grid of $N$ planes, layer separation $\Delta h$, ion mobility $K_{I}$, closed field $E_{c}$, average drift field within
grid $E_{o}$, drift length $L_{e}$, and electron drift velocity $v_{e}$, these times can be estimated as:

\begin{eqnarray*}
T_{o} & = & \frac{N\Delta h}{K_{I}E_{o}},\\
T_{c} & = & \alpha\frac{\Delta h}{K_{I}E_{c}},\\
T_{e} & = & \frac{L_{e}}{v_{e}}.
\end{eqnarray*}

The factor $\alpha$ in the collection time accounts for the fact that the field is not from parallel plates, but rather has low-field
regions in between the wires due to saddle points in the potential. Therefore $\alpha$ depends on the IBF threshold imposed. From the
above expressions, the live time can be written:
\begin{equation}
A=\frac{1-\frac{E_{o}K_{I}L_{e}}{N\Delta hv_{e}}}{1+\alpha\frac{E_{o}}{NE_{c}}}.\label{eq:1}
\end{equation}
This makes clear the dependence of the live time on various
detector parameters. The present simulations involve the following
parameter values, determined with the ALICE TPC in mind:

\begin{center}
\begin{tabular}{|c|c|c|c|c|c|}
\hline 
Param & Estimated value & $A\;\uparrow$ if & Physical reason & Constrained by\tabularnewline
\hline 
\hline 
$N$ & $4$ & $\uparrow$ & Longer $T_o$ & Transparency\tabularnewline
\hline 
$\Delta h$ & $3\;\text{mm}$ & $\uparrow$ & Longer $T_o$,$T_c$; fixed $T_{e}$ & Voltage; transparency\tabularnewline
\hline 
$E_{o}$ & $400-475\;\text{V/cm}$ & $\downarrow$ & Longer $T_o$ & Transparency\tabularnewline
\hline 
$E_{c}$ & $2000\;\text{V/cm}$ & $\uparrow$ & Faster collection & Voltage\tabularnewline
\hline 
$K_{I}$ & $4.8\;\text{cm}^2/\text{V}\cdot\text{s}$  & $\downarrow$ & Longer $T_o$,$T_c$; fixed $T_{e}$ & Gas choice\tabularnewline
\hline 
$v_{e}$ & $2.73\;\text{cm}/\mu\text{s}$ & $\uparrow$ & Smaller $T_{e}$ & Gas choice\tabularnewline
\hline 
$L_{e}$ & $250\;\text{cm}$ & $\downarrow$ & Smaller $T_{e}$ & Detector size\tabularnewline
\hline 
$\alpha$ & $1-4$ & $\downarrow$ & Longer collection time & IBF tolerance\tabularnewline
\hline 
$w$ & $2\;\text{mm}$ & $\downarrow$ & Smaller saddle area & Transparency\tabularnewline
\hline 
\end{tabular}
\par\end{center}

\subsubsection{Ion Collection}

\begin{figure}[!h]
\begin{centering}
%.334
\includegraphics[scale=0.65]{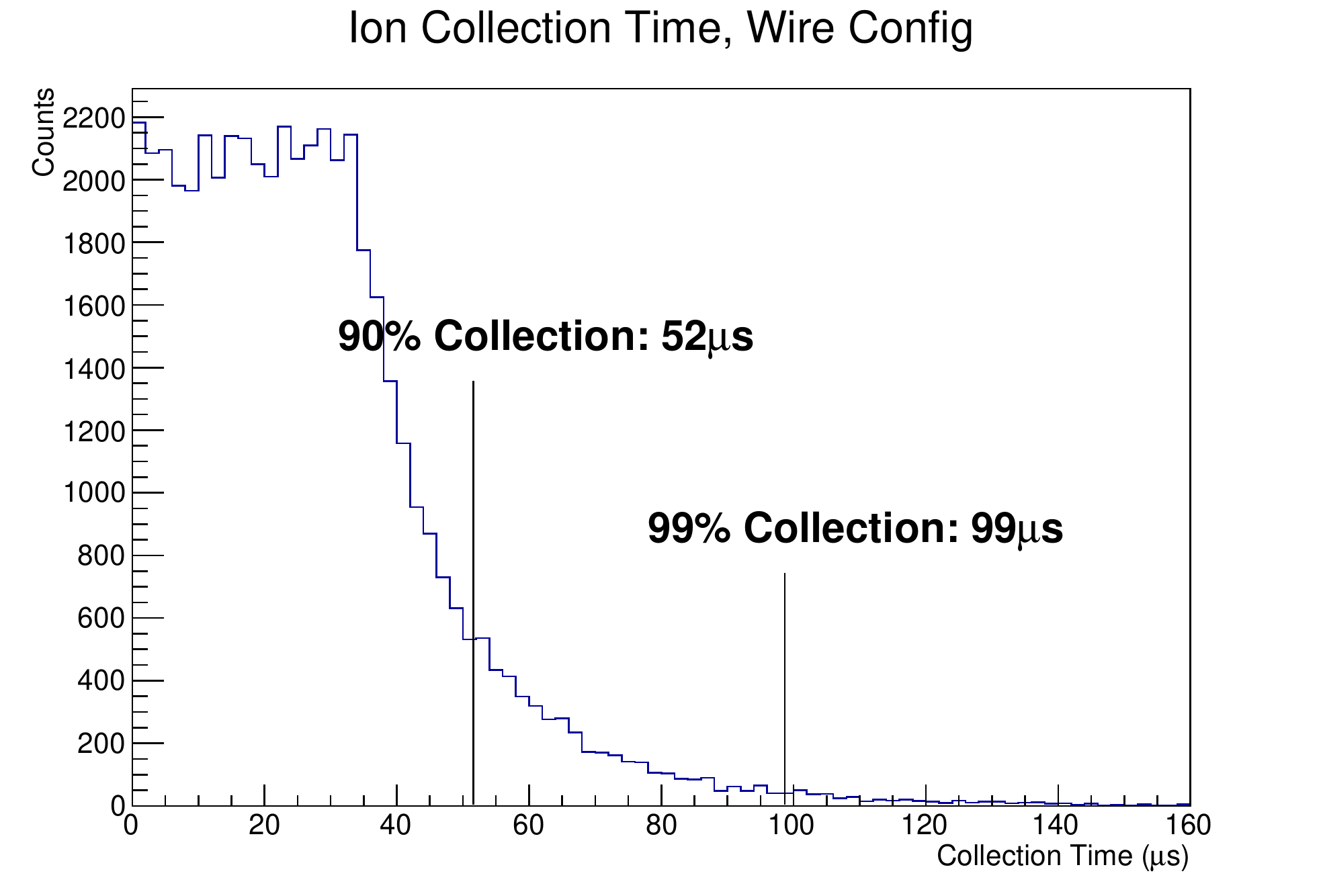}
\includegraphics[scale=0.65]{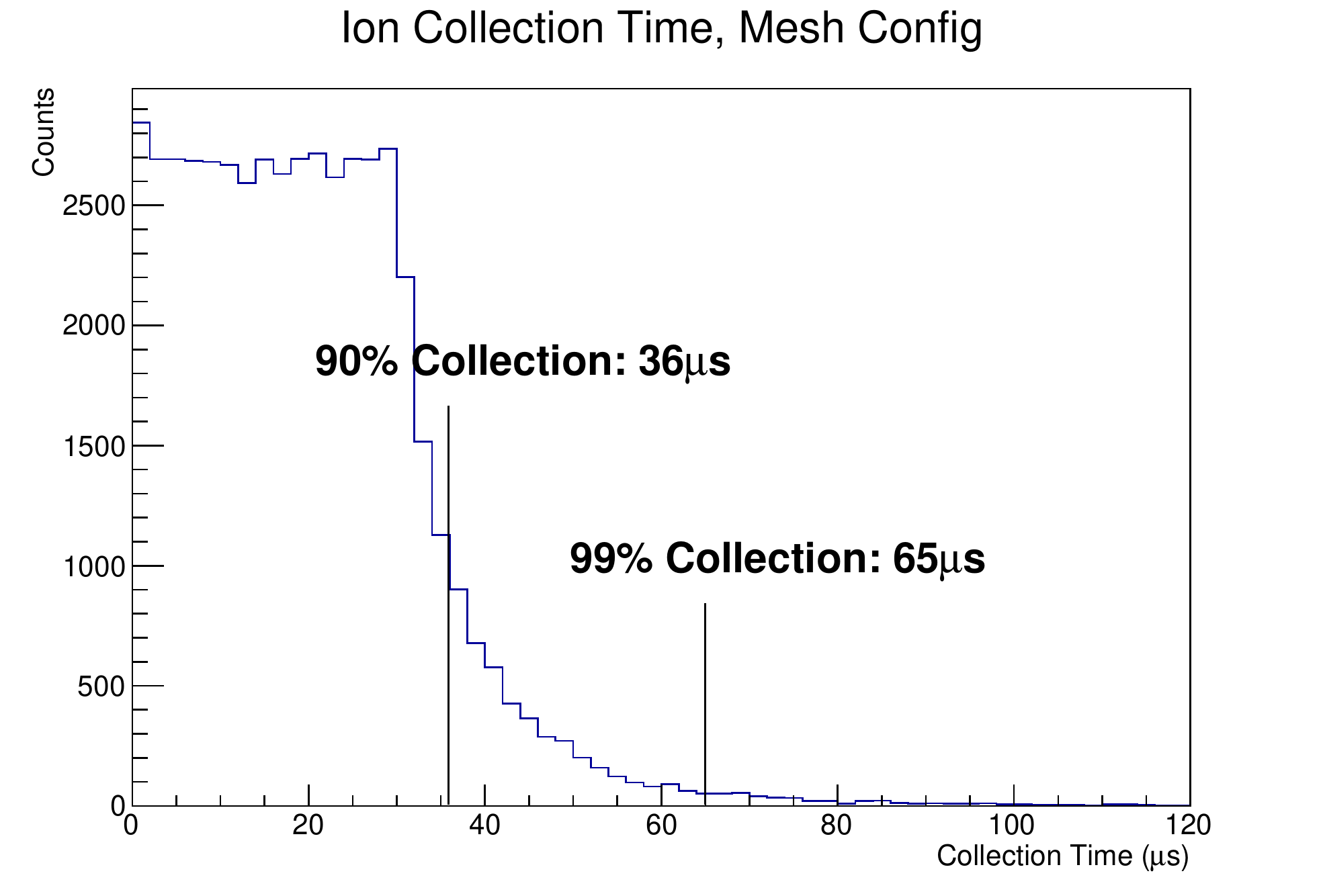}
\par\end{centering}
\caption{Histograms of $5\cdot10^{4}$ ion collection times in the ALICE configuration. The $90\%$ and
$99\%$ IBF thresholds are illustrated. Top: Wire-plane configuration.
Bottom: Mesh configuration. }
\label{fig:tcoll}
\end{figure}

To estimate the live time of the simulated configuration, the parameter $\alpha$ must be measured in simulation, or equivalently $T_{c}$.
I measure the ion collection time by randomly (uniformly) placing ions in the gating region, as one would expect for backdrifting ions,
and counting the time it takes to collect the ions in the closed configuration. Figure~\ref{fig:tcoll} shows the results. 

Note that a constant plateau out to $t=\frac{\Delta h}{K_{I}E_{c}}\approx31\;\mu\text{s}$
exists for both cases (as expected from a parallel plate solution), while the wire configuration has a significantly longer tail, due
to more low-field regions. 

Additionally, I introduced a magnetic field $B=0.5\;\text{T}$ in the mesh configuration, and repeated the ion collection. This causes
no change in collection times, as expected since the magnetic force on ions is negligible due to their slow drift velocities (and additionally,
the magnetic field would perturb ion trajectories not only into the low-field regions, but out of them as well).

\subsubsection{Live Time Estimates}

The measured collection times, in conjunction with the above table of parameters, yield the following live times, reported
for 90\% and 99\% IBF, and for transparencies corresponding to $\Delta E=0,20\;\text{V/cm}$, for both the wire-plane and 
mesh configurations:

\begin{center}
\begin{tabular}{|r|c|c|}
\hline 
 & Wire Configuration & Mesh Configuration\tabularnewline
\hline 
\hline 
$E_{o}=400\;\text{V/cm}$ & (80\% transparency) & (62\% transparency)\tabularnewline
\hline 
99\% IBF & 73\% & 77\%\tabularnewline
\hline 
90\% IBF & 78\% & 80\%\tabularnewline
\hline 
$E_{o}=450\;\text{V/cm}$ & (86\% transparency) & (68\% transparency)\tabularnewline
\hline 
99\% IBF & 70\% & 74\%\tabularnewline
\hline 
90\% IBF & 75\% & 78\%\tabularnewline
\hline 
\end{tabular}
\par\end{center}

\newpage

\section{Simulation Results -- STAR TPC}

Recalling equation (1), the corresponding table of values for STAR is estimated to be:

\begin{center}
\begin{tabular}{|c|c|c|}
\hline 
Param & Estimated value\tabularnewline
\hline 
\hline 
$N$ & $4$\tabularnewline
\hline 
$\Delta h$ & $3\;\text{mm}$\tabularnewline
\hline 
$E_{o}$ & $190\;\text{V/cm}$\tabularnewline
\hline 
$E_{c}$ & $2000\;\text{V/cm}$\tabularnewline
\hline 
$K_{I}$ & $1.6\;\text{cm}^2/\text{V}\cdot\text{s}$ \tabularnewline
\hline 
$v_{e}$ & $5.45\;\text{cm}/\mu\text{s}$\tabularnewline
\hline 
$L_{e}$ & $209\;\text{cm}$\tabularnewline
\hline 
$\alpha$ & $2$\tabularnewline
\hline 
$w$ & $2\;\text{mm}$\tabularnewline
\hline 
\end{tabular}
\par\end{center}

I take $E_{o}=190\;\text{V/cm}$, corresponding to a $140\;\text{V/cm}$
drift field plus $20\;\text{V/cm}$ per plane overvoltage. I then measure the electron
transparency with this overvoltage via simulation of 25,000
electrons in the wire-plane configuration to be:
\[
82.6\%.
\]
The statistical error is $\approx\sqrt{np(1-p)}/n\approx0.2\%$, but the dominant uncertainty is expected to come from the field map or
one of many other possible sources of error, which haven't been quantified. The transparency could be boosted at the expense of live time by
increasing the average gate field $E_{o}$. Also, small improvements $\sim1\%$ were observed in ALICE simulations when a magnetic field
was included, although this was not done here (the electrons in STAR are hotter than in ALICE, so the diffusion and $E\times B$ effects are both probably 
larger, and one would need to verify the outcome of their balance).  

\begin{figure}[!h]
\begin{centering}
\includegraphics[scale=0.65]{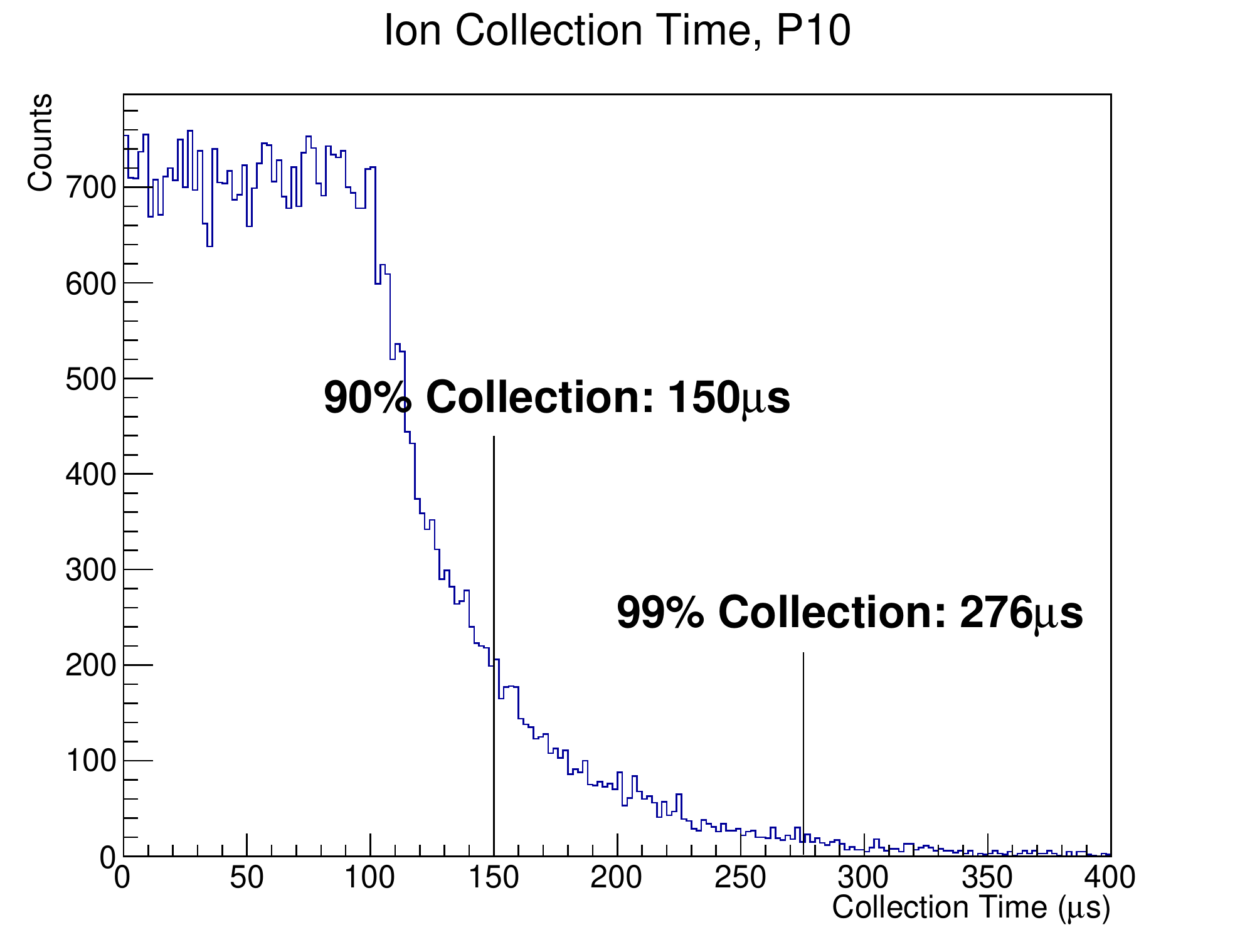}
\par\end{centering}
\caption{Histogram of $5\cdot10^{4}$ ion collection times in the STAR wire-plane configuration.
The 90\% and 99\% IBF thresholds are illustrated. }
\label{fig:tcollSTAR}
\end{figure}

I measure the parameter $\alpha$ by simulating the ion collection
time for 90\% and 99\% of backdrifting ions (Fig.~\ref{fig:tcollSTAR}). Recall that $\alpha=2$ means that the closed time is equal to twice that of a perfect
parallel plate.
I find 
\[
\alpha_{90\%}=1.6,\qquad\alpha_{99\%}=2.9.
\]
The live time for this set of parameters, with $82.6\%$ electron transparency, is then:
\[
A_{90\%}=95\%,\qquad A_{99\%}=93\%.
\]
Relative to ALICE, the smaller drift field and the smaller
ion mobility allow the open gate to be open longer (since ions drift back more slowly), and additionally the larger electron drift velocity
reduces the electron drift time per cycle $T_{e}$. Note also that in this scheme, the detector could operate continuously for up to
$T_{o}\sim4\;\text{ms}$ before the gate needs to be closed.

\section{Discussion}

The results above provide an early quantitative look at possible modified gating grid configurations. The specific results for live time and electron 
transparency should not be viewed as expected limits, but rather starting points from which optimization could begin.

One concrete conclusion, however, is that the mesh grid appears untenable. The idea of the mesh configuration is to increase the 
live time $A$ by decreasing $T_{c}$, at the expense of transparency. However, the
simulations suggest that the transparency cost is large for only a small improvement in live time. Further, it should be noted that the
wire configuration has an additional advantage in that it preserves momentum information along the direction of the wires, whereas the
mesh distorts momentum information in both directions. If one is determined to reduce $T_{c}$, an additional avenue to pursue is a dynamically
switched gating cycle, in which the saddle point ions are swept out of the low-field region. This could be accomplished by the closed
time consisting of two periods of $E_{c}\parallel E_{o}$ interspersed with a period of $E_{c}\perp E_{o}$.

A handful of additional parameters directly exhibit a tradeoff between live time and electron transparency: $N,\Delta h,\text{ and }E_{o}$. The idea
in designing a detector is to favor those variables that give maximal live time boost with minimal transparency loss. 

If electron transparency is a concern, one should increase $E_{o}$ as much as possible. To boost the live time, equation (\ref{eq:1}) 
suggests it is better to try to increase $T_{o}$ rather than decrease $T_{c}$. 
Further study of varying $N,\Delta h$, and $E_{o}$ should be undertaken. For example, in the ALICE configuration, plots of the final positions of electrons
show that transparency would decrease by approximately $3-5\%$ if another layer is added to the grid. This extra layer will cause $T_{o}\rightarrow\frac{5}{4}T_{o}$,
yielding live time improvements of approximately 5\%. Adding yet another layer would have an even smaller effect on transparency, and yield
a further boost in live time. Similar arguments can be made for increasing $\Delta h$, at the expense of longer collection time, perhaps worse
transparency, and larger voltage switches. This option may be particularly attractive for situations in which 90\% IBF suppression is acceptable.
The possibility of significantly increasing $E_{o}$ in concert with these approaches may be particularly appealing, and should be tested. 

To estimate the live time for configurations other than ALICE or STAR, one can use equation (\ref{eq:1}), upon choosing an $\alpha$ 
comparable to those for the ALICE and STAR results (for a given IBF suppression). Electron transparency estimates are more difficult, and require detailed simulation. 

\section{Conclusions}

The presented simulations suggest that a multi-layer extended gating grid may be a feasible option for reducing IBF in TPCs, depending on acceptable losses
of live time and electron transparency, and the TPC configuration. There remains significant room for optimization, and it is expected that results will continue 
to improve as they are adapted for particular applications. Experimental tests are also being pursued. 

\section{Acknowledgements}

Thanks to H. Wieman, J.W. Harris, R. Majka, and N. Smirnov for guidance and valuable discussions. Thanks also to the Garfield++ developers for
providing helpful documentation and useful examples from which to learn. 

This work was supported by the US Department of Energy under Grant DE-SC004168. 

This work was supported in part by the facilities and staff of the Yale University Faculty of Arts and Sciences High Performance Computing Center. 

%\section*{References}

\bibliography{Mulligan_MEGG}

\end{document}